# Bulk viscosity and deflationary universes

J. A. S. Lima*

*Departamento de Física Teórica e Experimental, Universidade Federal do Rio Grande do Norte, 59072, Natal, RN, Brasil*

R. Portugal

*Centro Brasileiro de Pesquisas Físicas, Rua Dr. Xavier Sigaud 150, Rio de Janeiro, 22290, RJ, Brasil*

I. Waga*

*Instituto de Física, Universidade Federal do Rio de Janeiro, Caixa Postal 68528, Rio de Janeiro, 21944, RJ, Brasil*

Abstract  We analyze the conditions that make possible the description of entropy generation in the new inflationary model by means of a near-equilibrium process. We show that there are situations in which the bulk viscosity cannot describe particle production during the coherent field oscillations phase.

The inflationary models of the early Universe are a very interesting solution to some shortcomings of the standard hot big-bang cosmology[1]. Usually these models contain two distinct phases. During the first one, called "slow rollover phase", the radius of the Universe increases by at least $10^{28}$ times while the temperature decreases by the same factor, maintaining constant the radiaton entropy. The second phase is in general called "coherent field oscillations phase " (CFOP) and during it both the radius of the Universe and temperature increase, generating radiation entropy[2,9]. This phase is essentially non-adiabatic, particles are produced through the damping of the coherent oscillations of the inflaton field by coupling to other fields and by it subsequently decay.

---

*J. A. S. Lima, R. Portugal and W. Waga*

Some time ago some authors[3,4] suggested that particle creation during or near the Planck era could be modeled classically by a bulk viscosity stress n. More recently Barrow[5] introduced this idea in the context of the inflationary models. In fact,, he was mainly interested in showing that when the strong energy condition is violated the "cosmic no hair" theorem fails. Working with a model first obtained by Murphy[6], in which the bulk viscosity coefficient is proportional to the energy density, he exhibited an exact solution of the Einstein equations that goes smoothly from an initial de Sitter phase to the Friedmann one. This is just what one would like to have occurring during the CFOP of the new inflationary models. So it is natural to ask if one can phenomenologically describe the particle production in this phase by a bulk viscosity.

It is well known that in an expanding Universe the effect of bulk viscosity is to diminish the effective pressure. It could in principle, simulate an equation of state of the form $p_{eff} = -\rho$, where $p_{eff} = p + \pi$, $\rho$ is the energy density and p the thermodynarnic pressure. So, it can drive an exponential inflation[7]. The bulk viscosity generates entropy also. In fact, it can be shown, in a thermodynamical setting, that in a spatially isotropic and homogeneous Universe the entropy production is non-vanishing if and only if the bulk viscosity is non-null.

In this paper, we analyze the conditions which allow one to describe the entropy production during the CFOP by a near-equilibrium process. To this end, we estimate the ratio $|\pi|/\rho$. If $|\pi|/\rho$ is greater than one, $\pi$ will be large enough to generate inflation during the CFOP, disturbing the expected evolution in the passage from the de Sitter stage to the Friedmann one. Only the models with $|\pi|/\rho \ll 1$ can, in principle, use the bulk viscosity to describe deflation.

Now, in a Friedmann-Robertson-Walker (FRW) background, by using the energy conçervation law

$$\dot\rho + 3H(\rho + p + \pi) = 0 \qquad (1)$$

where H is the Hubble parameter and the Gibbs law,

**Bulk viscosity and deflationary universes**

$$TdS = d(\rho R^3) + p dR^3 ,  \qquad (2)$$

where R is the scale factor of the Robertson-Walker metric, $S$ is the total entropy and $T$ is the temperature, we obtain the relation between the bulk viscosity and the entropy production

$$\pi = -\frac{T\dot{S}}{3HR^3} , \qquad (3)$$

where the overdot denotes time differentiation.

From non-equilibrium first-order thermodynamics[8], we know that $\pi = -3\varsigma H$, where $\varsigma$ is the bulk viscosity coefficient which is positive. So in an expanding Universe $\pi$ is negative.

From the Einstein equation

$$\ddot{R} = -\frac{4\pi}{3m_{\rm pl}^2}(\rho + 3p + 3\pi)R , \qquad (4)$$

where $m_{\rm pl} \approx 1.2 \times 10^{19}$ GeV is the Planck mass, we see that if $0 \leq p < \rho$, then to attain generalized inflation $\ddot{R} \geq 0$) it is sufficient for the absolute value of the bulk viscosity to be of the same order or greater than $p$.

From the hypothesis that the radiation entropy observed today was produced only in the inflationary period, the average entropy variation during the CFOP is given by

$$\dot{S} \cong T_{\rm rh}^3 R_{\rm rh}^3 \Gamma , \qquad (5)$$

where $\Gamma^{-1}$ is the mean lifetime of the inflaton field, $T_{\rm rh}$ is the final reheating temperature and $R_{\rm rh}$ the corresponding radius of the Universe.

We are considering a scalar field (inflaton field) $\phi$, whose Lagrangian density is given by

$$\mathcal{L} = -\frac{1}{2}\partial^\nu \phi \partial_\nu \phi - V(\phi) . \qquad (6)$$

The scalar field is homogeneous, and its equation of motion can be written as

*J. A. S. Lima, R. Portugal and W. Waga*

$$\frac{d}{dt}\left(\frac{1}{2}\dot{\phi}^2 + V\right) = -3H\dot{\phi}^2 \tag{7}$$

or equivalently

$$\frac{d}{dt}\rho_\phi = -3H(\rho_\phi + p_\phi) \tag{8}$$

where

$$\rho_\phi = \frac{1}{2}\dot{\phi}^2 + V(\phi) \quad \text{and} \quad p_\phi = \frac{1}{2}\dot{\phi}^2 - V(\phi) \ .$$

We shall assume that the frequency of the $\phi$ oscillations $\omega \cong \dot{\phi}/\phi)$ around the global minimum of the potential V, is always much greater than the expansion rate H. In this case, $\dot{\phi}^2$ can be replaced by its value averaged over an oscillation period[9]. By introducing the quantity $\gamma = <\dot{\phi}^2/\rho>$, the average of $\dot{\phi}^2/\rho$ over one cycle, eq. (8) can be integrated to give $\rho_\phi \propto R^{-3\gamma}$, where we have further assumed that $\gamma$ is constant.

Without invalidating the estimate we intend to do, we shall approximate the total energy density (p), by the coherent field energy density $(\rho_\phi)$. In fact, the radiation energy density $(\rho_{\rm rad})$ is negligible at the the beginning of the CFOP and is of the same order of $\rho_\phi$ at the end. So, at most, we will make an error by a factor of two. Then, in our approximation, the total energy density is given by

$$\rho \cong M^4 \left(\frac{R_b}{R}\right)^{3\gamma} \tag{9}$$

where $R_b$ is the radius of the Universe at the beginning of the CFOP and $M \cong 10^{14}$ GeV is the symmetry-breaking energy scale. By using eq. (9) and that at the end of the CFOP the energy density is $\rho \cong T_{\rm rh}^4$, we obtain

$$\frac{R_{\rm rh}}{R_b} \cong \left(\frac{M}{T_{\rm rh}}\right)^{4/3\gamma} \tag{10}$$

We are now ready to estimate the ratio $|\pi|/\rho$. By using eqs. (3), (5), (9) and (10), we find that

$$\frac{|\pi|}{\rho} \cong \left(\frac{R}{R_{\rm rh}}\right)^{3(\gamma-1)} \frac{T}{T_{\rm rh}} \frac{\Gamma}{H} \tag{11}$$

Bulk viscosity and *deflationary* universes

The value of $|\pi|/\rho$ depends on the kind of reheating. However, independent of it, at the beginning of the CFOP, when T is nearly zero we have $|\pi|/\rho << 1$.

Before we analyze eq. (11) in more detail we remark that the parameter $\gamma$ is associated to the form of the potential V. If, for example, one consider potentials of the form $V(\phi) = \alpha\phi^n$, where a, is a constant with dimensions of $(\text{mass})^{4-n}$, it is straightforward to show that

$$\gamma \equiv \left\langle \frac{\dot\phi^2}{\rho} \right\rangle = \frac{2n}{n+2} \qquad (12)$$

where < > means average over one cycle and it was assumed that on time scales $<< H^{-1}$, $\rho$ is constant[g]. For a renormalizable theory we must have $n \leq 4$ and consequently $\gamma \leq 4/3$. However, even if the Lagrangian (6) is an effective low-energy Lagrangian, eq. (12) shows us that $\gamma \leq 2$. An important and interesting case is that of a massive scalar field with $m^2 = V''(\phi = 0)$. In this case n = 2, $\gamma = 1$ and the coherent field behaves likes incoherent (dust) matter: $<p> = (\gamma-1)\rho = 0$, $\rho \propto R^{-3}$ and R $\propto t^{2/3}$.

Return now to eq. (11) and consider first the case when the scalar field decays faster than the characteristic expansion time, that is, H $<< \Gamma$ (good reheating). It is easy to see that at the end of the CFOP we have $|\pi|/\rho \cong \Gamma/H >> 1$. In this situation the maximum temperature coincides with the reheating temperature and although $|\pi|/\rho$ is small at the beginning of the CFOP, this ratio increases fast and becomes much greater than one. So, there would not occur deflation. Furthermore, the dominant energy condition

$$\rho + p + \pi \geq 0 \qquad (13)$$

would be violated, turning the picture physically inviable.

In the poor reheating case ($\Gamma << H$), the bulk viscosity does not need to increase as much. In fact, we can show this by calculating the ratio $|\phi|/\rho$ in two crucial instants. Consider first the instant t ⊢ $H^{-1}$, when the temperature achieves its maximum value[g]

$$T \cong \left(\frac{\Gamma}{H}\right)^{1/4} M \quad , \tag{14}$$

$R \cong R_b$ is minimum and $\pi$ is expected to be maximum. In this case, eq. (11) gives us that

$$\frac{|\pi|}{\rho} \cong \left(\frac{\Gamma}{H}\right)^{(11\gamma-8)/4\gamma} \ll 1 \quad , \text{if} \quad \gamma \geq 1 \quad . \tag{15}$$

For $t \cong \Gamma^{-1}$, that is, at the reheating instant, we have

$$\frac{|\pi|}{\rho} \cong \frac{\Gamma}{H} \ll 1 \quad . \tag{16}$$

So, in this case, the problems observed in the good reheating case would not appear.

In conclusion, we have analyzed under which conditions we can phenomenologically describe the particle production during the CFOP of the new inflationary model by a near-equilibriurn process.

Oiir analysis was performed by estimating the magnitude of the bulk viscosity necessary to generate the radiation entropy in the CFOP. In the good reheating case, we showed that the bulk viscosity should be so huge that would disturb the passage from the de Sitter to the Friedmann model. In this case, it is not possible to describe deflation using the bulk viscosity.

In the poor reheating case, it is possible, in principle, to describe deflation using the bulk viscosity, since its value will be much smaller than the total energy density during the CFOP.

We are gratcful to M. O. Calvão for useful comments. We would like to thank CAPES and FAPERJ for partial support.

### References

1. A. H. Guth, Phys. Rev. **D23**, 347 (1981); A. D. Linde, Phys. Lett. **B108**, 389 (1982); A. Albrecht and P. J. Steinhardt, Phys. Rev. Lett. 48, 1220 (1982).
2. A. Albrecht, P. J. Steinhardt, M. S. Turner and F. Wilczec, Phys. Rev. Lett. 48, **1437** (1982).

Bulk *viscosity* and *deflationary* universes

3. Ya. B. Zel'dovich, JETP Lett. 12, **307 (1970)**.
4. **B.** L. Hu, Phys. Lett. **A90,** 375 **(1982)**; **B. L.** Hu, Cosmology *of* the *Early* Universe, eds. L. Z. Fang and R. Ruffini (World Scientific, Singapore, **1984)**.
5. J. **D.** Barrow, Phys. Lett: **B180, 335 (1986)**.
6. **G. L. Murphy,** Phys. Rev. **D8, 4231 (1973)**.
7. J. A. **S.** Lima, R. Portugal and I. Waga, Phys. Rev. **D37, 2755 (1988)**.
8. **C.** Eckart, Phys. Rev. 58, **919 (1940)**; S. Weinberg, Ap. J. 168, **175 (1971)**; L. D. Landau and E. M. *Lifshitz,* Fluid *Mechanics* (Pergamon Press, Oxford, **1985)**.
9. **M. S.** Turner, Phys. Rev. D28, **1243 (1983)**.

Resumo

Nós analisainos as condições que possibilitam a descrição da produção de entropia no novo modelo inflacionário por um processo próximo ao equilíbrio termodinâmico. Nós mostramos que existem situações nas quais a viscosidade volumar não pode descrever a produção de partículas durante a fase de oscilações do campo coerente.